\documentclass[prd,twocolumn,twoside,preprintnumbers,superscriptaddress]{revtex4}

\usepackage{amsmath}
\usepackage{graphicx}
\usepackage{dcolumn}
\usepackage[hyperfootnotes=false]{hyperref}
\usepackage{xspace}
\usepackage{color}
\usepackage{url}

\newcommand{\bea}{\begin{eqnarray}}
\newcommand{\eea}{\end{eqnarray}}

\newcommand{\beq}{\begin{equation}}
\newcommand{\eeq}{\end{equation}}
\newcommand{\ec}{\end{center}}
\newcommand{\bc}{\begin{center}}
\newcommand{\tev}{{\rm TeV}}
\newcommand{\gev}{{\rm GeV}}

\newcommand{\cb}{{\cal{B}}}
\newcommand{\cl}{{\cal{L}}}

\definecolor{Red}{rgb}{1.,0.,0.}

\begin{document}
\preprint{\vbox{\hbox{LPT-Orsay-16-51}}}

\title{Leptoquark model to explain the $B$-physics anomalies, $R_K$ and $R_D$}

\author{Damir Be\v{c}irevi\'c}
\affiliation{Laboratoire de Physique Th\'eorique, CNRS, Univ.Paris-Sud, Universit\'e Paris-Saclay, 91405 Orsay, France}
\author{Svjetlana Fajfer}
\affiliation{Department of Physics, University of Ljubljana, Jadranska 19, 1000 Ljubljana, Slovenia}
\affiliation{Jo\v{z}ef \v{S}tefan Institute, Jamova 39, P.O.Box 3000, 1001 Ljubljana, Slovenia}
\author{Nejc Ko\v{s}nik}
\affiliation{Department of Physics, University of Ljubljana, Jadranska 19, 1000 Ljubljana, Slovenia}
\affiliation{Jo\v{z}ef \v{S}tefan Institute, Jamova 39, P.O.Box 3000, 1001 Ljubljana, Slovenia}
\author{Olcyr Sumensari}
\affiliation{Laboratoire de Physique Th\'eorique, CNRS, Univ.Paris-Sud, Universit\'e Paris-Saclay, 91405 Orsay, France}
\affiliation{Instituto de F\'isica, Universidade de S\~ao Paulo, C.P. 66.318, 05315-970 S\~ao Paulo, Brazil.}

\begin{abstract}
We show that a model with a scalar leptoquark of hypercharge $Y=1/6$ which includes the light right-handed neutrinos, can successfully describe both of the $B$-physics anomalies, $R_K^{\rm exp} < R_K^{\rm SM}$ and 
$R_D^{\rm exp} > R_D^{\rm SM}$. We discuss the corresponding low energy effective theory and, after using the known experimental data as constraints, we show that the model is viable and that it offers several predictions
which can be tested experimentally. 
\end{abstract}

\maketitle

\section{Introduction}
\label{intro}
Even though the LHC results so far did not unveil the new physics (NP) particles, the $B$-physics experiments at LHCb and at the $B$-factories pointed 
at a very intriguing effects of lepton flavor universality violation (LFUV). 
More specifically, the LHCb Collaboration measured the partial branching fractions of $B\to K\ell^+\ell^-$ which, integrated over $q^2 \in
    [1,6] \ \gev^2$, resulted in Ref.~\cite{Aaij:2014ora}
\begin{align}\label{exp:RK}
R_K = \frac{ \cb( B \to K \mu \mu)}{\cb( B \to K e e)} = 0.745 \pm^{0.090}_{0.074} \pm 0.036 \,,
\end{align}
$2.6 \sigma$ below the Standard Model (SM) prediction, $R_K^{\rm SM} =1.00(1)$~\cite{Hiller:2003js}. Another intriguing indication of LFUV was unveiled in the processes 
mediated by the charged currents and measured at the $B$-factories where it was found~\cite{Lees:2012xj},
\begin{align}
R_D = \left.\frac{ \cb( B \to D \tau \nu_\tau)}{\cb( B \to D
  l \nu_l)}\right|_{l \in \{e,\mu\}} \!\! = 0.41 \pm 0.05 \,,
\end{align}
$1.9 \sigma$ larger than the SM prediction, $R_D^{\rm SM}=0.286\pm 0.012$, obtained by solely relying on the lattice QCD data for both the vector and the scalar form factors, 
recently presented in~\cite{MILC}. That result is corroborated by the experimentally established $R_{D^\ast}=0.317\pm 0.017$, also confirmed by LHCb~\cite{Aaij:2015yra}, 
which appears to be $3.3 \sigma$ larger than predicted, $R_{D^\ast}^{\rm SM}=0.252\pm 0.003$~\cite{Fajfer:2012jt}. Note, however, that for the theoretical estimate of 
$R_{D^\ast}^{\rm SM}$ the form factors were extracted from the angular distribution of $d\Gamma(B\to D^\ast \mu \nu_\mu)/dq^2$, up to a normalization, and the validity 
of leading order heavy quark effective theory has been assumed in evaluating the pseudoscalar form factor. The lattice QCD result for the full set of $B\to D^\ast$ form factors is not available. 


Several models have been proposed in order to simultaneously describe LFUV in $R_K$ and $R_{D^{(\ast)}}$. By using a set of gauge invariant NP operators made of left-handed fermions 
the authors usually assume that only the coupling to one generation in the interaction basis is non-zero so that the LFUV comes from the misalignment between the interaction and mass bases. 
In Ref.~\cite{Bhattacharya:2014wla}  it was assumed that a satisfactory description can be made by setting only the coupling to the third generation to be non-zero. A similar route has been followed 
by Ref.~\cite{Alonso:2015sja} where the couplings to other generations are kept non-zero but suppressed by factors $\propto m_{e,\mu}^2/m_\tau^2$. Further contribution in this direction has been 
made in Ref.~\cite{Calibbi:2015kma}, as well as in Ref.~\cite{Feruglio:2016gvd}, where it has been argued that the effects of the  renormalization group running from the NP scale to the electroweak 
symmetry breaking scale can generate the effects of LFUV particularly significant in the decays of $\tau$-lepton. Another model building option consists in adding a $SU(2)_L$ triplet of massive gauge 
bosons that couple to one generation of fermions~\cite{Greljo:2015mma}, an option which leads to tensions with direct searches at the LHC.  Assuming that the LFUV comes from the difference 
of the lepton numbers, then $L_\mu-L_\tau$ can be promoted into a gauge symmetry which, when enriched by one generation of vector-like leptons, results in a substantial modifications of 
the $\tau\to 3\mu$ and $h\to \mu\mu$ decay rates~\cite{Altmannshofer:2016oaq}. Finally, to accommodate both $R_K$ and $R_{D^{(\ast )}}$ the scenarios with a hypothetical light leptoquark (LQ) states 
have been proposed. While the scenarios with vector LQ's are the easiest ones~\cite{Fajfer:2015ycq} they become problematic when computing the loop corrections unless the vector LQ's are promoted 
into the ``{\sl light}" gauge bosons ${\cal O}(1\ \tev)$, in which case one runs into contradiction as such gauge bosons are supposed to be associated with a gauge group relevant to the scales of grand unification. 
Otherwise the loop corrections in a theory with a light vector LQ are UV-cutoff dependent unless the UV completion is explicitly specified.  
Concerning the light scalar LQ scenarios, instead, they do not exhibit such a problem but in their minimal form they are suitable to either describe $R_K$~\cite{Becirevic:2015asa} or $R_{D^{(\ast )}}$~\cite{Dorsner:2013tla}, 
but not both. In this paper we argue that a minimal extension of the scalar LQ with the hypercharge $Y=1/6$ can lead to a simultaneous description of both $R_K$ and $R_{D}$. 
Finally, we should also mention that in Ref.~\cite{Bauer:2015knc} it has been argued that a model with the simplest $SU(2)_L$-singlet scalar LQ one can accommodate $R_K$ through a loop correction 
and $R_{D^{(\ast )}}$ via the tree-level LQ contribution. That scenario has been challenged in Ref.~\cite{US} where it was shown that a simultaneous description of $R_K$ and $R_{D^{(\ast)}}$ is not realistic 
and that accommodating the experimental value of $R_K$ would imply serious phenomenological problems elsewhere.

In the following we first describe our model in Sec.~\ref{sec:model}. The expressions for quantities used as constraints are given in Sec.~\ref{scan} where we perform 
the scan of parameters and show that the model accommodates both $B$-physics anomalies. Several significant predictions are presented in Sec.~\ref{sec:predictions} and we conclude in Sec.~\ref{concl}. 

\section{Leptoquark Model\label{sec:model}}
The Yukawa Lagrangian for a theory with the LQ state $\Delta$ which carries the quantum numbers $(3,2)_{1/6}$ of the SM gauge group, $(SU(3)_c,SU(2)_L)_Y$, in the interaction basis reads,
\bea
\cl_{\Delta} &=& {\overline d_R^\prime Y_L (\widetilde \Delta)^\dagger
  L^\prime +  \overline Q^\prime Y_R \Delta \nu_R^\prime +
  \mathrm{ h.c.} ,}
\label{eq:L1}
\eea
where the standard notation has been used, with $L$ and $Q$ being the left-handed doublet of leptons and quarks respectively, combined with the right-handed (RH) singlet fermions 
and with the $SU(2)_L$-doublet of LQ fields $ \Delta$, where $\widetilde \Delta = i\sigma_2 \Delta^\ast$. The primed fermion fields ($\psi^\prime$) are related to the unprimed ones through rotations, 
$(\psi^i)^\prime_{L,R} = U^{i \dag}_{L,R} \psi^i_{L,R}$, so that after
taking $Y_{L} \to {U_{R}^{d} Y_{L} U_L^{\ell\dag}}$ and
$Y_{R} \to {U_{L}^{d} Y_{R} U_R^{\nu\dag}}$, one recognizes the Pontecorvo-Maki-Nakagawa-Sakata 
$U_{\rm PMNS}= {U_L^\ell  U_L^{\nu \dag}}$, and the Cabibbo-Kobayashi-Maskawa $V_{\rm CKM} = U_L^u U_L^{d\dag}$ matrices, and the above Lagrangian in the fermion mass eigenbasis becomes,
\begin{align}\label{eq:eff2}
&\cl_{\Delta} = \overline d_R \left( Y_L U_{\rm PMNS}\right) \nu_L\Delta^{(-1/3)} - \overline d_R Y_L\ell_L\Delta^{(2/3)}  \cr
& + \overline u_L \left(V_{\rm CKM} Y_R \right) \nu_R\Delta^{(2/3)} + \overline d_L Y_R\nu_R\Delta^{(-1/3)} + \mathrm{ h.c.}, 
\end{align}
where the superscript in $\Delta^{(Q)}$ denotes the electric charge eigenstates of the LQ doublet, $Q=Y+T_3$, which we assume to be degenerate in mass ($T_3$ being the weak isospin). 
The couplings $Y_{L,R}$ are the $3\times 3$ matrices. The crucial difference between Eq.~(\ref{eq:eff2}) and the model discussed in Ref.~\cite{Becirevic:2015asa} is the presence of the second 
line in Eq.~(\ref{eq:eff2}). In other words, besides the doublet of light scalar LQ states with hypercharge $Y=1/6$, in this model we also have the light RH neutrinos the mass of which 
is assumed to be very small with respect to the hadronic mass scale, and in the following we will neglect it. We consider neutrinos to be Dirac particles, even though this issue is immaterial in the limit 
of $m_\nu \to 0$. Since the neutrinos are considered as massless it is legitimate to take $U_{\rm PMNS}$ to be the unit matrix.

The above Yukawa Lagrangian is the essential ingredient of the full model which also comprises the kinetic and mass terms of the LQ field. 
Our working assumption is that $m_\Delta \simeq 1$~TeV, and since we are working with the low energy processes it is more convenient to work in a low energy effective effective theory, 
obtained by integrating out the heavy propagating $\Delta$. We first
focus onto the terms relevant to $b\to s\ell\ell$ and $b\to
c\ell\overline\nu$ transitions. For the first one we obtain
\begin{align}\label{eq:Lbsmm}
&\cl^{d_k\to d_i\ell\ell}_{\rm eff} = \frac{1}{m_\Delta^2} Y_L^{ij} Y_L^{\ast kl} \  \overline d_i P_L\ell_j\  \overline \ell_l P_R d_k +\mathrm{ h.c.}\cr
& \quad = - \frac{Y_L^{ij}  Y_L^{\ast kl}  }{2 m_\Delta^2}\ \overline d_i\gamma_\mu P_R d_k \ \overline \ell_l \gamma^\mu P_L\ell_j +\mathrm{ h.c.} \,,
\end{align}
where the second line is obtained by applying the Fierz identity. $P_{L/R}= (1\mp\gamma_5)/2$, as usual. 
For the charged current process, instead, we have, 
\begin{align}\label{eq:Leff2}
\cl^{d\to u\ell\overline \nu}_{\rm eff} = & \frac{(V_{\rm CKM} Y_R)^{ij}  Y_L^{\ast kl}  }{2 m_\Delta^2}   \left[ 
\overline u_i P_R d_k \ \overline \ell_l P_R\nu_j \right.\cr
&\left. \qquad +\frac{1}{4} \overline u_i \sigma_{\mu\nu}P_Rd_k\  \overline \ell_l \sigma^{\mu\nu} P_R\nu_j \right] + \mathrm{ h.c.}\,,
\end{align}
which means that the NP contribution to the semileptonic decays (and to $b\to c\ell \bar\nu$ in particular) arising from this model comes with the non-zero RH Yukawa couplings. 
Furthermore, in the low energy effective theory one also generates the
process $c\to u\nu\bar \nu$ which is not phenomenologically interesting in
the massless neutrino limit.  
Another significant contribution generated by this model is the one related to $b\to s\nu\bar\nu$ transition, namely,   
\begin{align}\label{eq:LeffNUNU}
&\cl^{d_k\to d_i\nu\bar \nu}_{\rm eff} =  -\sum_{\alpha=L,R}{  Y_\alpha^{ij}Y_\alpha^{\ast kl} \over 2 m_\Delta^2 } 
\bar d_i\gamma^\mu (1-P_\alpha ) d_k\bar \nu_l\gamma_\mu P_\alpha \nu_j  \cr
 & - { Y_L^{ij}Y_R^{\ast kl} \over 2 m_\Delta^2 } \left[ \bar d_i P_Ld_k \bar \nu_l P_L\nu_j +\frac{1}{4} \bar d_i \sigma_{\mu\nu} P_Ld_k \bar \nu_l \sigma^{\mu\nu} P_L\nu_j \right]  + \mathrm{ h.c.},
\end{align}
which will be used in the phenomenological discussion below.

\section{Constraints on Yukawa Couplings\label{scan}}

In this work, for simplicity, we will take the couplings to the first generation to be zero in order to avoid the potential problems with the atomic parity violation experiments~\cite{Dorsner:2016wpm}, and we will 
assume the following structure of the matrices of Yukawa couplings:
\[
Y_{L,R} = \left( \begin{matrix}
  0 & 0 & 0\\
  0 & Y_{L,R}^{s \mu} & Y_{L,R}^{s \tau}\\
  0 & Y_{L,R}^{b \mu} & Y_{L,R}^{b \tau}
\end{matrix}\right), 
\] 
\[
V_{\rm CKM}Y_{R} = \left( \begin{matrix}
  0 & V_{us} Y_{R}^{s\mu} + V_{ub}  Y_{R}^{b\mu} & V_{us}  Y_{R}^{s\tau} + V_{ub} Y_{R}^{b\tau}\\
  0 & V_{cs} Y_{R}^{s\mu} + V_{cb}  Y_{R}^{b\mu} & V_{cs}  Y_{R}^{s\tau}  + V_{cb} Y_{R}^{b\tau}\\
  0 & V_{ts} Y_{R}^{s\mu} + V_{tb}  Y_{R}^{b\mu} & V_{ts}  Y_{R}^{s\tau}  + V_{tb} Y_{R}^{b\tau}
\end{matrix}\right)\,.
\]
The product $V_{\rm CKM}Y_{R}$ is explicitly written in order to emphasize the fact that even if the couplings that involve the first generation of quarks/leptons are zero, the NP contributions to the leptonic 
and semileptonic decays of kaons ($s\to u$) or $B$-mesons ($b\to u$), driven by the Lagrangian~(\ref{eq:Leff2}), are not absent. The values of the couplings  $Y_{L,R}^{ij}$, which we take to be real, 
are varied within the perturbativity limits, $\vert (Y_{L,R})_{ij}\vert \leq 4\pi$, and are subjects to many phenomenological constraints of which the following ones are found to be particularly efficient:\\

{\bf 1.} As in Ref.~\cite{Becirevic:2015asa} we use the experimentally established $\cb(B_s\to\mu\mu)$~\cite{CMS:2014xfa} and $\cb(B\to K\mu\mu)$ in the large $q^2$-bin~\cite{Aaij:2014pli}, 
and we combine them with the lattice QCD values for $f_{B_s}$ and for the $B\to K$ form factors~\cite{FLAG}, to extract $(C_9^{\mu\mu})^\prime=- (C_{10}^{\mu\mu})^\prime \in (-0.48,-0.08)$. 
This result is equivalent to constraining the combination $Y_{L}^{b \mu} Y_{L}^{s \mu}/m_\Delta^2$, which then leads to $R_K=0.88(8)$, consistent with the experimental value found by LHCb, cf. Eq.~(\ref{exp:RK}). 
Since the NP contribution is mediated by the RH currents, this model predicts $R_{K^\ast} =1.11(9)$, i.e. in contrast with the models in which the NP gives rise to the non-primed Wilson coefficients in which case 
$R_{K^\ast} < 1$, as discussed in Ref.~\cite{Hiller:2014ula}.\\

{\bf 2.} Another important constraint on $Y_{L,R}$ stems from the $B_s- \overline  B_s$ mixing. We compute $\Delta m_{B_s}$ in our model, divide it by its well known 
SM expression and obtain,
\begin{align}\label{eq:BBbar}
&R_{B_s}={\Delta m_{B_s}\over \Delta m_{B_s}^{\rm SM}} = 1 + {\eta_1\over 16 G_F^2 m_W^2 (V_{tb}V^\ast_{ts})^2\eta_BS_0(x_t) m_\Delta^2}\times \cr
&\qquad \left[ (Y_L Y_L^\dag)_{bs}^2 + \frac{1}{2} (Y_R Y_R^\dag)_{bs}^2 -   \eta_{41} \frac{3}{2}  (Y_L Y_L^\dag)_{bs} (Y_R Y_R^\dag)_{bs}\times \right.\cr
&\left.
\qquad\, \times \left( {m_{B_s}\over m_b(m_b) + m_s(m_b)}\right)^2 {B_4(m_b)\over B_1(m_b)}
\right], 
\end{align}
where we use the (standard) notation for $\Delta m_{B_s}^{\rm SM}$, $\eta_1 = 0.82(1)$ and $\eta_{41} = 4.4(1)$ account for the QCD evolution from $\mu\simeq 1$~TeV down to $\mu=m_b$. 
After combining the lattice QCD values for bag parameters $B_{1,4}$~{\cite{FLAG,ETMC}, with the experimental $R_{B_s}^{\rm exp}=1.02(10)$, 
we obtain a rather stringent constraint on the couplings shown in the brackets of Eq.~(\ref{eq:BBbar}).\\

{\bf 3.} The upper experimental limit on the lepton flavor violating decay, $\cb(\tau\to \phi\mu)$~\cite{PDG}, provides an efficient constraint on $Y_L^{s\tau}$ via
\begin{align}
\cb(\tau\to \phi \mu)=\frac{f_\phi^2 m_\tau^3}{512 \pi  \Gamma_\tau}\left|\frac{Y_L^{s\tau}Y_L^{ s\mu \ast}}{m_\Delta^2} \right|^2 (1-x)(1+x-2x^2),
\end{align}
where $x=m_\phi^2/m_\tau^2$, $f_\phi=241(18)$~MeV~\cite{cd2}, and we omitted writing the terms $\propto m_\mu^2/m_\tau^2$.\\

{\bf 4.} Also useful are the constraints coming from the
(semi-)leptonic meson decays. We find it more convenient to work with
the following effective Lagrangian, 
 \bea\label{eq:Leff3}
 \cl_{\rm eff}&=& - 2\sqrt{2} G_F V_{ud}\left[ \overline u\gamma_\mu P_L d\ \overline \ell \gamma_\mu P_L \nu + g_S\,\overline u P_R d\ \overline \ell P_R \nu\right. \cr
  &&\qquad \left.+ g_T\, \overline u \sigma_{\mu\nu} P_R d\ \overline \ell \sigma^{\mu\nu}P_R \nu \right] +\mathrm{ h.c.},
 \eea
where $u$/$d$ stands for a generic up-/down-type quark, while $g_{S,T}\equiv g_{S,T}^{d\to u \ell\bar \nu}$ are the NP couplings introduced in such a way that in the limit in which they vanish one retrieves the usual SM Fermi theory. 
Using this Lagrangian one can easily compute the decay rates for various leptonic processes.  
For example, 
\begin{equation}
  \label{eq:lept}
  \begin{split}
\Gamma(D_s\to \ell \bar \nu)&= \frac{G_F^2}{8\pi m_{D_s}^3}   \vert V_{cs}\vert^2 f_{D_s}^2 (m_{D_s}^2-m_\ell^2)^2 m_\ell^2  \cr
&\qquad \times \left[1+\vert g_S\vert^2{ m_{D_s}^4\over  m_\ell^2 (m_c+m_s)^2}\right],   
  \end{split}
\end{equation}
where we used  $\langle 0\vert \overline c\gamma_\mu\gamma_5 s\vert D_s(p)\rangle = i f_{D_s}p_\mu$.  From the matching of decay rates obtained with~(\ref{eq:Leff3}) and with~(\ref{eq:Leff2}) we 
get 
\begin{align}
g_S^{c\to s\ell_i  \nu_j} = { \left( V_{\rm CKM} Y_R\right)^{cj}   Y_L^{si \ast }\over 4\sqrt{2} G_F V_{cb}m_\Delta^2}\,, 
\end{align}
at the scale of the mass of the LQ, $\mu = m_\Delta \simeq 1$~TeV, which is then via the QCD running related to the low scale value as 
$g_S(1\ \tev)\approx 2\times g_S(\mu=m_b)\approx 2.7\times   g_S(\mu=2\ \gev)$. Notice that in this model, in order to have a non-vanishing NP contribution to the 
processes driven by the charged currents in the SM, the relevant RH coupling(s) should be non-zero. 
Since there is no interference term between the SM and the NP terms, we can consider the effective coupling to be
\bea
|g_S^{c\to s\mu \nu}|^2 =  |g_S^{c\to s\mu \nu_\mu}|^2 + |g_S^{c\to s\mu \nu_\tau}|^2, 
\eea
and {\it mutatis mutandis} for the other leptonic decays. In other words, in this theory  
the flavor state of neutrino is not specified but can be both $\nu_\mu$ and $\nu_\tau$, i.e. 
$\cb(D_s\to \tau \bar \nu)= \cb(D_s\to \tau \bar \nu_\tau)+\cb(D_s\to \tau \bar \nu_\mu)$. 
By adequately using the above expression and combining it with the experimentally established $\cb(K\to \mu\nu)$, $\cb(\tau\to K\nu)$, $\cb(D_s\to \mu \nu)$, $\cb(D_s\to \tau \nu)$, $\cb(B\to \tau \nu)$~\cite{PDG}, together 
with the relevant decay constants given in Ref.~\cite{FLAG}, 
we obtain the valuable constraints on various Yukawa couplings. \\

{\bf 5.} Since the non-zero NP coupling to muons is essential to describe $R_K^{\rm exp}<1$, while keeping such a coupling to electrons set to zero, it is now important to make sure that 
\bea
R_D^{\mu/e}= \frac{ \cb( B \to D \mu \nu)}{\cb( B \to D e\nu)} ,
\eea
remains small. The relevant expression for the differential decay rate, obtained by using the Lagrangian~(\ref{eq:Leff3}), reads
\begin{align}\label{eq:SLrate}
\frac{d\cb}{dq^2} &= \cb_0 \vert V_{cb}\vert^2 |f_+(q^2)|^2 \left\{ c_+^\ell(q^2) + \vert g_T|^2 c_T^\ell(q^2) \left|\frac{f_T(q^2)}{f_+(q^2)}\right|^2\right.\cr
&\left.  + \left( 1+\vert g_S\vert^2{ q^4\over  m_\ell^2 (m_b-m_c)^2}\right) c_0^\ell(q^2) \left|\frac{f_0(q^2)}{f_+(q^2)}\right|^2\right\},\end{align}
where $\cb_0= G_F^2\tau_B/(192\pi^3m_B^3)$, and the coefficient functions are given by
\begin{align}
&c_+^\ell(q^2)=\lambda^{3/2}\left[1-\frac{3}{2}\frac{m_\ell^2}{q^2}+\frac{1}{2}\left(\frac{m_\ell^2}{q^2}\right)^3\right] \,,\cr
&c_T^\ell(q^2)=\lambda^{3/2}{2 q^2\over (m_B+m_D)^2}\left[1-3\left( \frac{m_\ell^2}{q^2}\right)^2 +2\left(\frac{m_\ell^2}{q^2}\right)^3\right] \,,\cr
&c_0^\ell(q^2)=m_\ell^2\lambda^{1/2} \frac{3}{2}\frac{m_B^4}{q^2} \left(1- \frac{m_\ell^2}{q^2}\right)^2 \left(1- \frac{m_D^2}{m_B^2}\right)^2 ,
\end{align}
with $\lambda\equiv \lambda(q^2)=[q^2-(m_B+m_D)^2][q^2-(m_B- m_D)^2]$. The form factors $f_{0,+,T}(q^2)$ in~(\ref{eq:SLrate}) are defined as usual, 
\begin{equation}
  \begin{split}
    \hspace{-.5em} &\langle D(k)|\bar c \gamma_\mu b|B(p)\rangle = (p+k)_\mu f_+(q^2)+q_\mu f_-(q^2),\cr
    \hspace{-.5em} &\langle D(k)|\bar c \sigma_{\mu\nu} b|B(p)\rangle = -i(p_\mu k_\nu - k_\mu p_\nu)  {2 f_T(q^2)\over (m_B+m_D) },
  \end{split}
\end{equation}
and $f_-(q^2) = [f_0(q^2)- f_+(q^2)]\times (m_B^2-m_D^2)/q^2$.
Using the form factors from Ref.~\cite{MILC} and requiring $R_D^{\mu/e} < 1.05$, we will obtain quite a powerful constraint on our couplings,
\bea\label{eq:gST}
g_S^{b\to c\ell_i\bar \nu_j}=4 \ g_T^{b\to c\ell_i\bar \nu_j} = { \left( V_{\rm CKM} Y_R\right)^{cj}  Y_L^{bi \ast }\over 4\sqrt{2} G_F V_{cb}m_\Delta^2}\,, 
\eea 
where the tensor coupling scales as $g_T(1\ \tev)\approx 0.78 \times g_T(\mu=m_b)\approx 0.7\times   g_T(\mu=2\ \gev)$, and the tensor form factor is taken from~\cite{tens}. \\

{\bf 6.} Finally, the experimental upper limit on $\cb(B\to K \nu\bar \nu)$~\cite{PDG} turns out to be an important constraint too. 
The relevant expression for this process computed in the SM, extended by the effective Lagrangian~(\ref{eq:LeffNUNU}), is
\begin{align}\label{eq:BKnunu}
&{d\cb \over dq^2}=  {
\alpha_{\rm em}^2 G_F^2 |V_{tb}V^\ast_{ts}|^2 \tau_B\over 768 \pi^5 m_B^3} \lambda^{3/2} | f^{B\to K}_+(q^2)|^2 \times  \cr
& \left\{ 3|C_L^{\rm SM}|^2 - {\mathrm{Re}}\left[ {C_L^{\rm SM} ( Y_L Y_L^\dag )_{sb} \over 2 {\cal N}    m_\Delta^2 }\right] +{\cal O}\left(\frac{1}{m_\Delta^4}\right)\right\},
\end{align}
where ${\cal N}= \alpha_{\rm em}G_F V_{tb}V^\ast_{ts}/(\sqrt{2} \pi)$, $C_L^{\rm SM}=-6.38(6)$~\cite{brod}, and we do not show the terms $\propto 1/m_\Delta^4$.\\

With all of the above ingredients in hands we are now able to constrain the Yukawa couplings $Y_{L,R}^{ij}$, which are then used to determine the values of $g_{S}^{b\to c\mu\bar \nu}$ 
and $g_{S}^{b\to c\tau\bar \nu}$, while the corresponding tensor couplings are obtained by using Eq.~(\ref{eq:gST}) at the scale $\mu= m_\Delta$.  After inserting those final couplings 
into Eq.~(\ref{eq:SLrate}) we can compute $R_D = \cb(B\to D\tau  \nu)/\cb(B\to Dl \nu)$. The result is shown in Fig.~\ref{fig:1} where we see that 
with all of the constraints discussed above, our model not only gives $R_K=0.88(8)$ compatible with the experimental finding, but we are also able to find the points which are compatible 
with $R_D^{\rm exp}$ to $1 \sigma$. 
\begin{figure}[t]
\begin{center}
\includegraphics[width=0.43\textwidth]{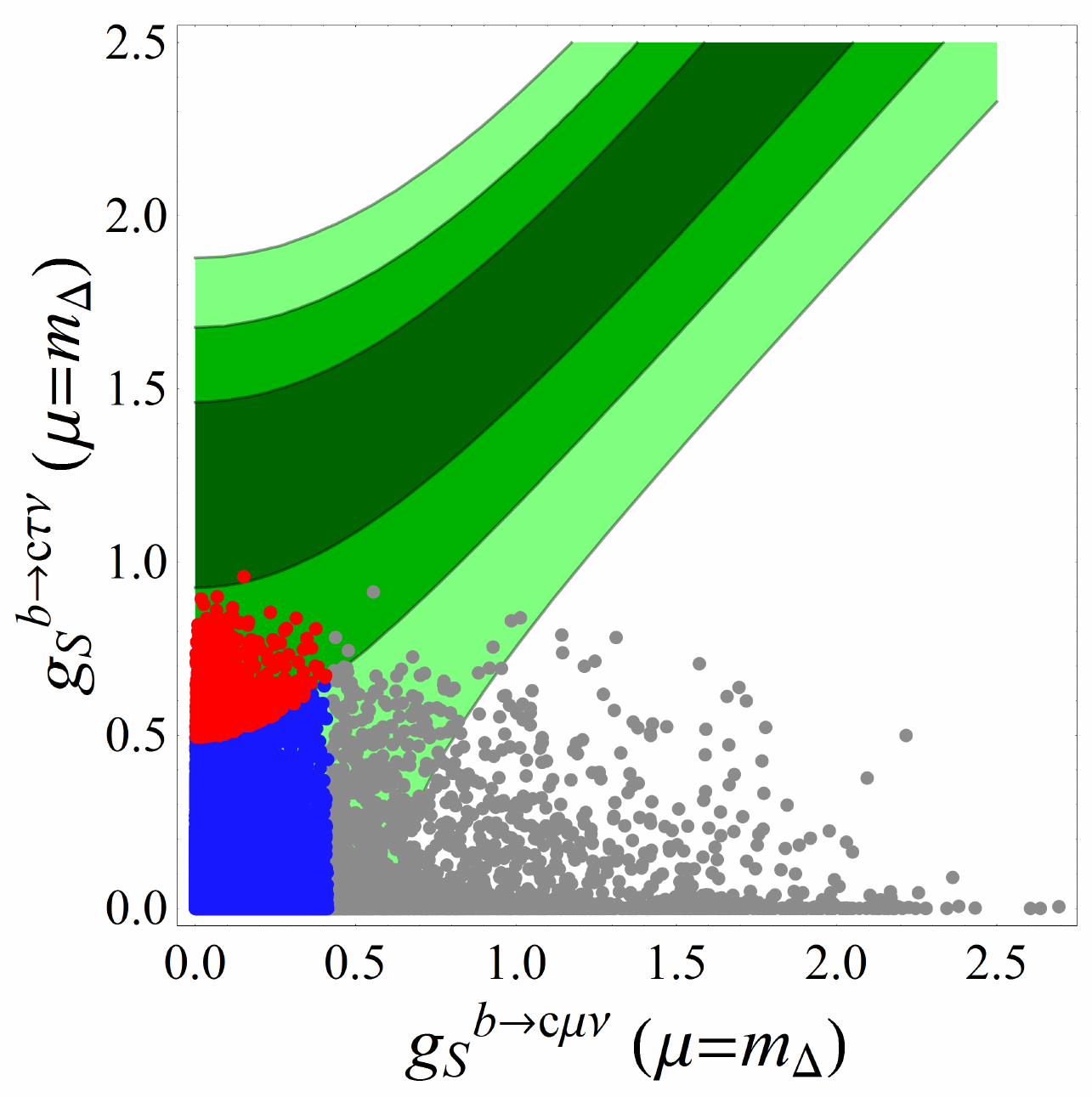}
\end{center}
\caption{The ensemble of points (all colors combined) correspond to our model after applying all the constraints on Yukawa couplings discussed in this Section except for the constraint {\bf 5}. They are shown in the plane 
$g_{S}^{b\to c\tau\bar \nu}$ Vs. $g_{S}^{b\to c\mu\bar \nu}$, against the green regions which represent $R_D$ at $1$-, $2$- and $3$-$\sigma$.  
Red and blue points are selected after imposing the condition {\bf 5.}. Finally the red points alone indicate the compatibility with $R_D$ to $2\sigma$. \label{fig:1}}
\end{figure}
In other words, the model we propose here can satisfactorily accommodate both $B$-physics anomalies, $R_K$ and $R_D$. We should reiterate that we focused on $R_D$ because all of the form factors have been 
computed on the lattice, and we do not need to rely on the experimental information about the normalization and shapes of the form factors, which is not the case with $R_{D^\ast}$. Using the experimental information 
about the form factors would be inappropriate in our case since we claim that both the couplings to $\tau$ and to $\mu$ are modified. We should say that by using the model form factors, such as those from Ref.~\cite{melikhov}, we indeed obtain 
that $R_{D^\ast}>R_{D^\ast}^{\rm SM}$ and in a good ballpark with respect to the experimental results, but we prefer not to quote those results until the lattice QCD determination of the full set of form factors becomes available.

The structure of Yukawa couplings from the constraints listed above is such 
that $Y_{L,R}^{s\mu}$ and $Y_{L,R}^{s\tau}$ are small, while $Y_{L,R}^{b\mu}$ and $Y_{L,R}^{b\tau}$ can be large and are correlated in such a way that $Y_{L}^{b\mu}$ and $Y_{L,R}^{b\tau}$ are large for small values of $Y_{R}^{b\mu}$, 
but diminish in size with the increase of $Y_{R}^{b\mu}$. We checked that $\Gamma(K\to \mu\nu)/\Gamma(K\to e\nu)$ remains intact, i.e. at its SM value. We also checked that our model is consistent with the direct LQ searches~\cite{DS}, and that varying $m_\Delta \in (0.7, 1)$~TeV, 
leaves our conclusions unchanged. Notice also that in the scalar LQ model with $Y=1/6$ the enhancement of $\cb(\tau \to \mu\gamma)$ and $(g-2)_\mu$ are highly suppressed and experimentally indistinguishable from their SM predictions~\cite{US,Dorsner:2016wpm}.

\section{Predictions\label{sec:predictions}}

\begin{figure*}[t]
\vspace*{-6mm}
\begin{center}
\begin{tabular}{ccc}
\includegraphics[width=0.33\textwidth]{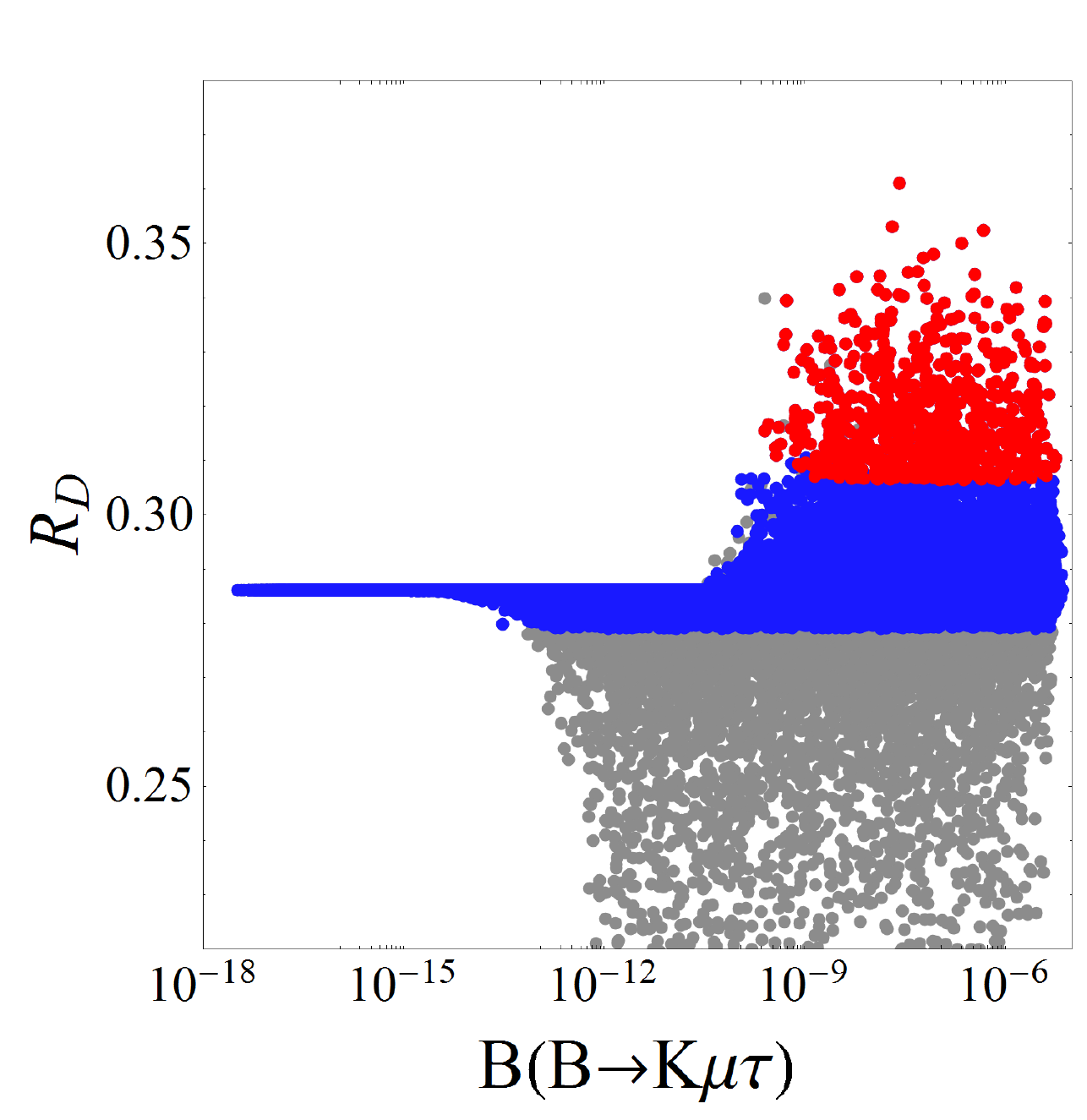}
\includegraphics[width=0.33\textwidth]{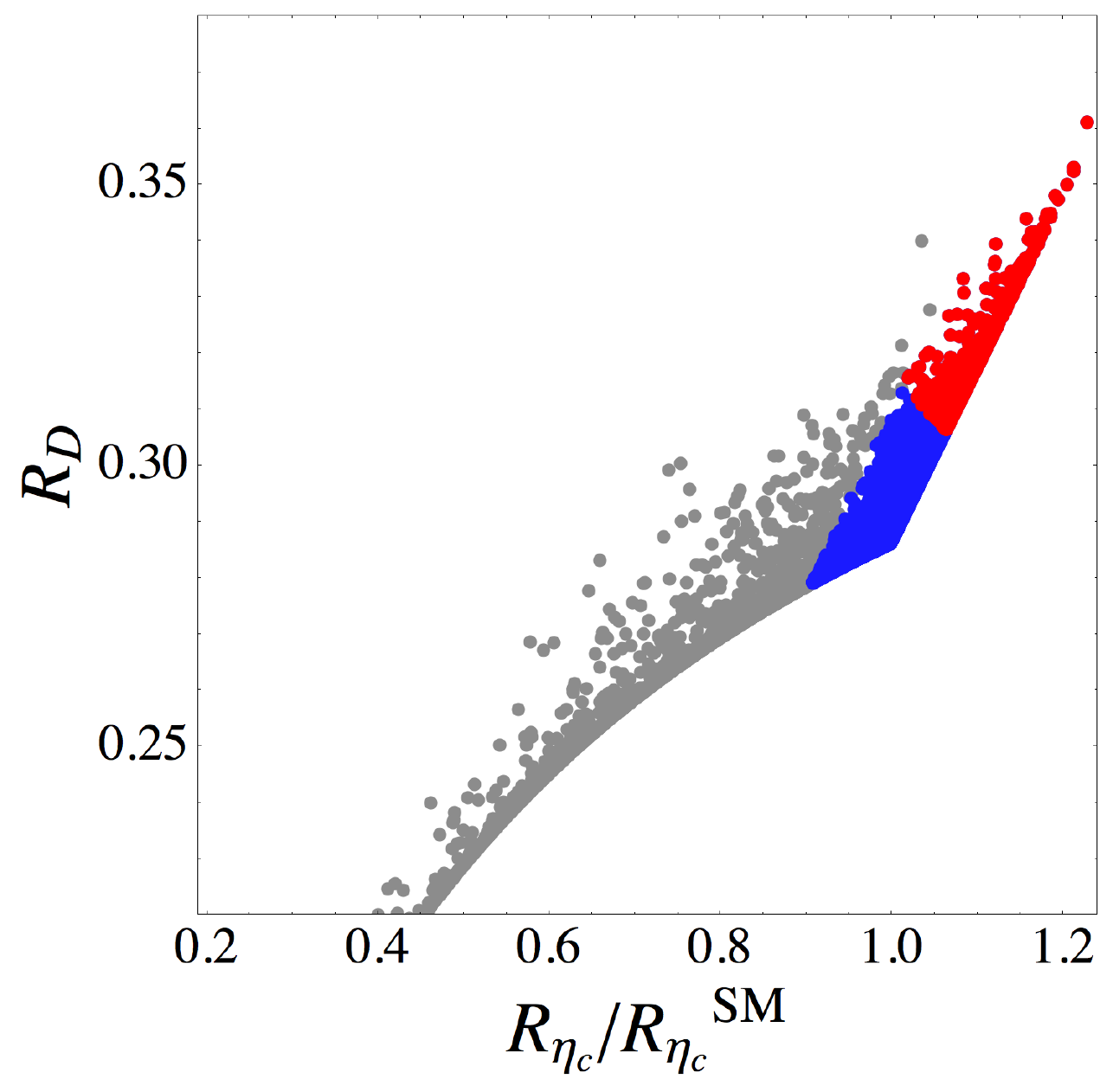}
\includegraphics[width=0.33\textwidth]{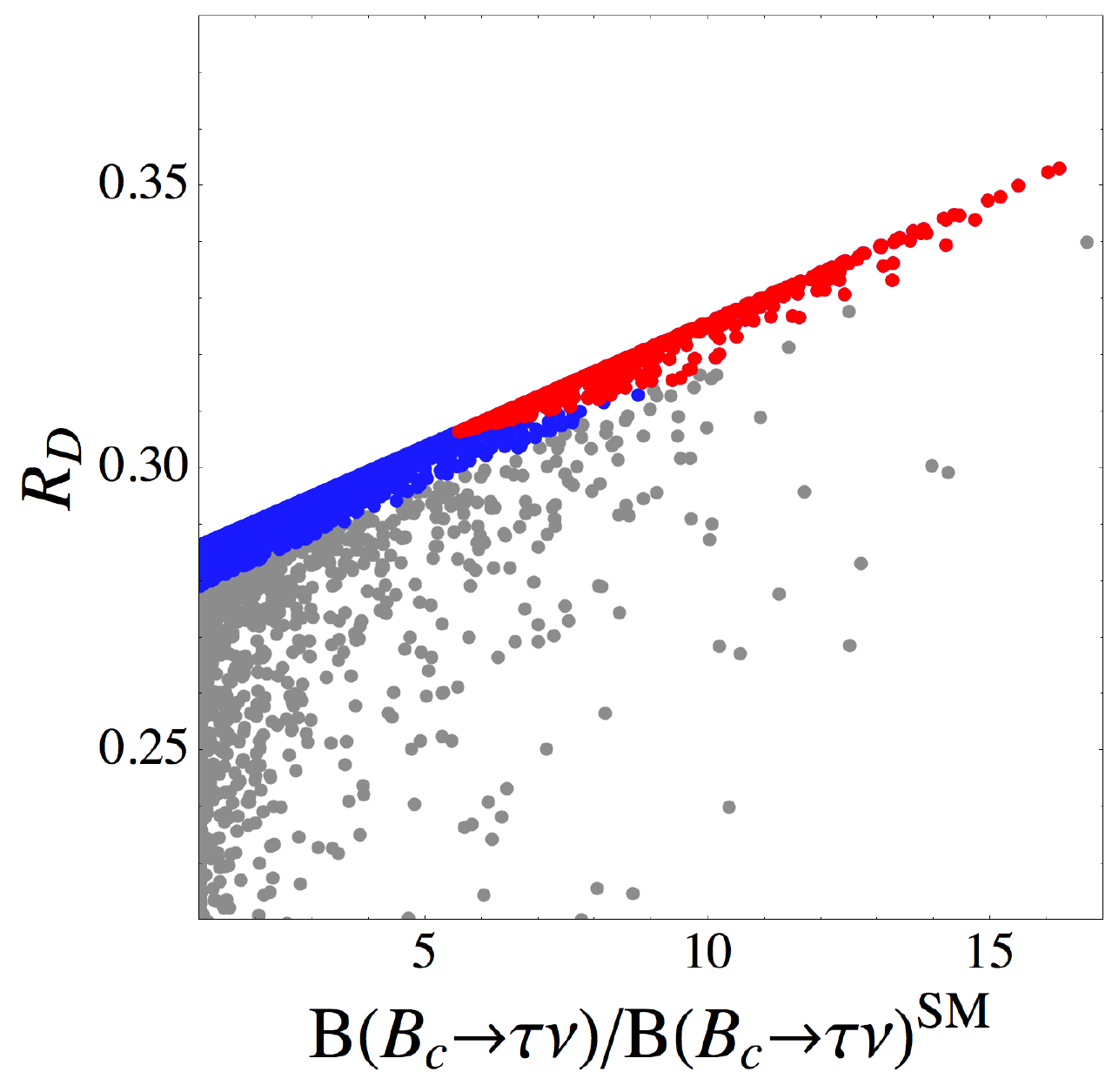}
\end{tabular}
\end{center}
\caption{Colors of the points are the same as in Fig.~\ref{fig:1}: the blue points are obtained by subjecting the Yukawa couplings of our model to the constraints discussed in Sec.~\ref{scan}, and the red ones are selected from the blue ones after requiring the compatibility with 
$R_D^{\rm exp}$ to $2\sigma$. We plot our predictions for three selected quantities: limits on the LFV decay mode, i.e. its $\cb (B\to K\mu\tau)$, the ratio between $R_{\eta_c}=\cb(B_c\to \eta_c\tau\bar \nu)/(B_c\to \eta_c l\bar \nu)$ predicted by our model and its SM value, and 
a similar ratio of $\cb(B_c\to \tau\bar \nu)$ which appears to be strictly larger than its SM estimate. \label{fig:2}}
\end{figure*}

With the Yukawa couplings constrained in a way discussed in the previous Section, we could show that we are able to accommodate both $R_K^{\rm exp}$ and $R_D^{\rm exp}$, and in this Section we discuss several predictions that we obtain. 
Besides an important and verifiable prediction, $R_{K^\ast}=1.11(9)$, which is a peculiarity of our model~\cite{Becirevic:2015asa}, we also find the following:
\begin{itemize}
\item The value of $\cb(B_s\to \tau\tau)$ can be both larger and smaller than the SM one because the Wilson coefficient, $C_{10}^\prime = \pi Y_L^{b\tau }Y_L^{s\tau}/(2 \sqrt{2}G_FV_{tb}V_{ts}^\ast \alpha_{\rm em}m_\Delta^2)$, can be negative and positive respectively. We get
\bea
0 \leq {\cb(B_s\to \tau\tau)/\cb(B_s\to \tau\tau)^{\rm SM}}  < 33.
\eea
\item Using the expressions presented in Ref.~\cite{ourLFV}, we also computed the lepton flavor violating decay $\cb (B\to K\mu\tau)$ and found that 
\bea
2.1\times 10^{-10} \leq {\cb(B\to K\mu\tau)}  \leq 6.7\times 10^{-6},
\eea
which is shown in Fig.~\ref{fig:2}. Notice that the similar LFV modes are easily inferred from the bounds given above, by using $\cb(B_s\to \tau\mu) \approx  0.9\times \cb(B\to K\mu\tau)$, and 
$ \cb(B\to K^\ast\mu\tau) \approx  1.8\times \cb(B\to K\mu\tau)$, cf. Ref.~\cite{ourLFV}. This result is similar to what has been obtained in Ref.~\cite{Calibbi:2015kma}, except that they have $R_{K^\ast}<1$.
\item 
Just like $R_D > R_D^{\rm SM}$, we find that the ratio $R_{\eta_c}=\cb(B_c\to \eta_c\tau \nu)/(B_c\to \eta_c l \nu)$ can be larger than its SM value. Using the recent $B_c\to \eta_c$ decay form factor values computed on the lattice~\cite{cd3} 
and the results for $g_S^{b\to c\ell_i\bar \nu_j}$ discussed above, we obtain 
\bea
1.02 \leq {R_{\eta_c}/R_{\eta_c}^{\rm SM}}  \leq 1.21,
\eea
which is also plotted in Fig.~\ref{fig:2}.
\item A very interesting feature of this model is not only that the different leptonic decays of $B_c$ are modified differently, but the fact that $\cb(B_c\to \tau \nu)$ we obtain is strictly larger than the SM value. 
We get
\bea
5.5 \leq {\cb(B_c\to \tau \nu)/\cb(B_c\to \tau \nu)^{\rm SM}}  \leq 16.1,
\eea
which offers another possibility to experimentally test the validity of our model.
On the other hand, the value of $\cb(B_c\to \mu \nu)$ that we obtain can be either equal to its SM value or enhanced by up to a few orders of magnitude.  
\item  We computed $\cb(t\to b\tau  \nu)$, both in the SM and in our model and found, 
\begin{align}
&{d\cb(t\to b\tau  \nu)^{\rm SM}\over dq^2} = |V_{tb}|^2 \lambda^{1/2} {(m_t^2-q^2)(m_t^2+2q^2)\over 3072\pi^3 m_t^3 \Gamma_t} \times \cr
&\hspace*{4cm} {g_W^4\over (m_W^2-q^2)^2+m_W^2\Gamma_W^2},\cr
&{d\cb(t\to b\tau  \nu) \over dq^2}   = {d\cb(t\to b\tau  \nu)^{\rm SM}\over dq^2}  \times \cr
&\qquad \left[ 1+ {Y_L^{b\tau} (V_{\rm CKM}Y_R)^{t\ell} \over |V_{tb}|^2}{(m_W^2-q^2)^2+m_W^2\Gamma_W^2\over m_\Delta^4}\right],
\end{align}
where $g_W^4=32 m_W^4 G_F^2$, and $q^2\in [m_\tau^2, (m_t-m_b)^2]$. By using the constraints from Sec.~\ref{scan} and for $m_\Delta \in (0.7,1)$~TeV, we find that  
\bea
{ \cb(t\to b\tau  \nu) -\cb(t\to b\tau  \nu)^{\rm SM}\over \cb(t\to b\tau  \nu)^{\rm SM}} \leq 5 \times 10^{-3},
\eea
i.e. indistinguishable from the SM value even if the experimental uncertainty is improved by several orders of magnitude. Notice that the current experimental error is $30\%$~\cite{PDG}.
\end{itemize}

\section{Conclusions\label{concl}}

In this paper we propose a model which can accommodate both $B$-physics anomalies that hint on the LFUV, namely $R_K^{\rm exp}< R_K^{\rm SM}$ and $R_D^{\rm exp}> R_D^{\rm SM}$. 
The model is a scenario with the doublet of mass degenerate light scalar leptoquarks, with hypercharge $Y=1/6$ and the mass around $1$~TeV, which was already known to be viable in obtaining $R_K^{\rm exp}< R_K^{\rm SM}$.
The novelty is that we include the light RH neutrinos, that we consider to be massless, which entail new operators and give rise to the matrix of RH Yukawa couplings. The suitable products of left-handed and RH couplings, if non-zero, can modify 
the leptonic and semileptonic decay rates. We then show that by using the available experimental information as constraints we were able to accommodate $R_D^{\rm exp}$. Since the model modifies the decays to both muons and to $\tau$-leptons (but not to electrons), 
we could not provide the numerical assessment of the similar $R_{D^\ast}$ except that we indeed get $R_{D^\ast}^{\rm exp}> R_{D^\ast}^{\rm SM}$. 

Another interesting feature is that this model provides at least two experimentally verifiable predictions: (i) We find that $R_{K^\ast} =1.11(9)$, and (ii) $\cb(B_c\to \tau \nu)$ is $5\div 16$ times larger than the SM prediction. 
Other predictions are listed in the body of the paper. Our results for the exclusive LFV $b\to s\mu\tau$ modes are similar to what is obtained in other scenarios, namely that their branching fractions can be up to ${\cal O}(10^{-6})$. Finally, we suggest that it could be interesting to check 
for the LFUV effects in $R_{\eta_c}=\cb(B_c\to \eta_c\tau \nu)/(B_c\to \eta_c l \nu)$, which in our model is larger than predicted in the SM. 

{\it Acknowledgment}: {\small  
This project has received funding from the European Union's Horizon 2020 research and innovation program under the Marie Sklodowska-Curie grant agreement No. 674896.
S.F. and N.K. acknowledge support of the Slovenian Research Agency (research core funding No. P1-0035).}


\begin{thebibliography}{99}

\bibitem{Aaij:2014ora}
  R.~Aaij {\it et al.} [LHCb Collaboration],
  Phys.\ Rev.\ Lett.\  {\bf 113} (2014) 151601
  [arXiv:1406.6482].

\bibitem{Hiller:2003js}
  G.~Hiller and F.~Kruger,
  Phys.\ Rev.\ D {\bf 69} (2004) 074020
  [hep-ph/0310219];
M.~Bordone {\it et al.},
  arXiv:1605.07633.

\bibitem{Lees:2012xj}
  J.~P.~Lees {\it et al.} [BaBar Collaboration],
  Phys.\ Rev.\ Lett.\  {\bf 109} (2012) 101802
  [arXiv:1205.5442];
 M.~Huschle {\it et al.} [Belle Collaboration],
  Phys.\ Rev.\ D {\bf 92} (2015) no.7,  072014
  [arXiv:1507.03233];
A.~Abdesselam {\it et al.} [Belle Collaboration],
  arXiv:1603.06711.

\bibitem{MILC}
  J.~A.~Bailey {\it et al.} [MILC Collaboration],
  Phys.\ Rev.\ D {\bf 92} (2015) no.3,  034506
  [arXiv:1503.07237].


\bibitem{Aaij:2015yra}
  R.~Aaij {\it et al.} [LHCb Collaboration],
  Phys.\ Rev.\ Lett.\  {\bf 115} (2015) no.11,  111803
   Addendum: [Phys.\ Rev.\ Lett.\  {\bf 115} (2015) no.15,  159901]
  [arXiv:1506.08614].


\bibitem{Fajfer:2012jt}
  S.~Fajfer, J.~F.~Kamenik, I.~Nisandzic and J.~Zupan,
  Phys.\ Rev.\ Lett.\  {\bf 109} (2012) 161801
  [arXiv:1206.1872].



\bibitem{Bhattacharya:2014wla}
  B.~Bhattacharya, {\it et al.},
  Phys.\ Lett.\ B {\bf 742} (2015) 370
  [arXiv:1412.7164].



\bibitem{Alonso:2015sja}
  R.~Alonso, B.~Grinstein and J.~Martin Camalich,
  JHEP {\bf 1510} (2015) 184
  [arXiv:1505.05164].

\bibitem{Calibbi:2015kma}
  L.~Calibbi, A.~Crivellin and T.~Ota,
  Phys.\ Rev.\ Lett.\  {\bf 115} (2015) 181801
  [arXiv:1506.02661].


\bibitem{Feruglio:2016gvd}
  F.~Feruglio, P.~Paradisi and A.~Pattori,
  arXiv:1606.00524.


\bibitem{Greljo:2015mma}
  A.~Greljo, G.~Isidori and D.~Marzocca,
  JHEP {\bf 1507} (2015) 142
  [arXiv:1506.01705];
 S.~M.~Boucenna {\it et al.},
  Phys.\ Lett.\ B {\bf 760} (2016) 214
  [arXiv:1604.03088], and
  arXiv:1608.01349.


\bibitem{Altmannshofer:2016oaq}
  W.~Altmannshofer, M.~Carena and A.~Crivellin,
  arXiv:1604.08221.



\bibitem{Fajfer:2015ycq}
  S.~Fajfer and N.~Ko\v{s}nik,
  Phys.\ Lett.\ B {\bf 755} (2016) 270
  [arXiv:1511.06024];
R.~Barbieri {\it et al.},
 Eur.\ Phys.\ J.\ C {\bf 76} (2016) no.2,  67
 [arXiv:1512.01560];



\bibitem{Becirevic:2015asa}
  D.~Be\v{c}irevi\'c, S.~Fajfer and N.~Ko\v{s}nik,
  Phys.\ Rev.\ D {\bf 92} (2015) no.1,  014016
  [arXiv:1503.09024].

\bibitem{Dorsner:2013tla}
  I.~Dor\v{s}ner, S.~Fajfer, N.~Ko\v{s}nik and I.~Ni\v{s}and\v{z}i\'c,
  JHEP {\bf 1311} (2013) 084
  [arXiv:1306.6493].



\bibitem{Bauer:2015knc}
  M.~Bauer and M.~Neubert,
  Phys.\ Rev.\ Lett.\  {\bf 116} (2016) no.14,  141802
  [arXiv:1511.01900].


\bibitem{US}   D.~Bečirević, N.~Košnik, O.~Sumensari and R.~Zukanovich Funchal,
  arXiv:1608.07583 [hep-ph].

\bibitem{Dorsner:2016wpm}
  I.~Dor\v{s}ner, S.~Fajfer, A.~Greljo, J.~F.~Kamenik and N.~Ko\v{s}nik,
  Phys.\ Rept.\  {\bf 641} (2016) 1
  [arXiv:1603.04993].

\bibitem{CMS:2014xfa} 
  V.~Khachatryan {\it et al.} [CMS and LHCb Collaborations],
  Nature {\bf 522}, 68 (2015)
  [arXiv:1411.4413].


\bibitem{Aaij:2014pli} 
  R.~Aaij {\it et al.} [LHCb Collaboration],
  JHEP {\bf 1406}, 133 (2014)
  [arXiv:1403.8044].

\bibitem{FLAG}
  S.~Aoki {\it et al.},
  arXiv:1607.00299.


\bibitem{Hiller:2014ula}
  G.~Hiller and M.~Schmaltz,
  JHEP {\bf 1502} (2015) 055
  [arXiv:1411.4773].

\bibitem{ETMC}
  N.~Carrasco {\it et al.} [ETM Collaboration],
  JHEP {\bf 1403} (2014) 016
  [arXiv:1308.1851].

\bibitem{PDG} 
  K.~A.~Olive {\it et al.} [Particle Data Group Collaboration],
  Chin.\ Phys.\ C {\bf 38}, 090001 (2014).

\bibitem{cd2}
  G.~C.~Donald {\it et al.} [HPQCD Collaboration],
  Phys.\ Rev.\ D {\bf 90} (2014) no.7,  074506
  [arXiv:1311.6669].

\bibitem{tens}
  M.~Atoui, V.~Mor\'enas, D.~Be\v{c}irevi\'c and F.~Sanfilippo,
  Eur.\ Phys.\ J.\ C {\bf 74} (2014) no.5,  2861
  [arXiv:1310.5238].

\bibitem{brod}
  J.~Brod, M.~Gorbahn and E.~Stamou,
  Phys.\ Rev.\ D {\bf 83} (2011) 034030
  [arXiv:1009.0947].

\bibitem{melikhov}
  D.~Melikhov and B.~Stech,
  Phys.\ Rev.\ D {\bf 62} (2000) 014006
  [hep-ph/0001113].

\bibitem{DS}
 S.~I.~Cooper [CMS Collaboration] and S.~Viel [Atlas Collaboration], talks presented at ICHEP-2016.
 
\bibitem{ourLFV}
  D.~Be\v{c}irevi\'c, O.~Sumensari and R.~Zukanovich Funchal,
  Eur.\ Phys.\ J.\ C {\bf 76} (2016) no.3,  134
  [arXiv:1602.00881].


\bibitem{cd3}
  A.~Lytle {\it et al.},
  arXiv:1605.05645.

\end{thebibliography}
\end{document}